\documentclass[twocolumn,showpacs,preprintnumbers,amsmath,amssymb]{revtex4}

\usepackage{graphicx}
\usepackage{dcolumn}
\usepackage{bm}

\begin{document}

\title{
 Secular variability, geodetic precession and moment of inertia
of binary pulsars}
\author{B.P. Gong}

\affiliation{ Department of Astronomy, Nanjing University, Nanjing
210093,PR.China}

\email{bpgong@nju.edu.cn}


\begin{abstract}
More and more binary pulsars show significant secular variations,
in which the measured projected semi-major axis, $\dot{x}^{obs}$,
and the first derivative of orbital period, $\dot{P}_{b}^{obs}$,
 are several order of magnitude larger than the prediction of
general relativity (GR). This paper shows that the geodetic
precession induced orbital effects can explain both $\dot{x}$ and
$\dot{P}_{b}$ measured in binary pulsars. Moreover, by this model
we can automatically estimate the magnitude of the spin angular
momenta of the pulsar and its companion star, and therefore the
moment of inertia ( $10^{44}$g~cm$^2$ to $10^{45}$g~cm$^2$) of
pulsar of binary pulsar systems, which agrees well with
theoretical predictions. In other words, the contamination
(residuals represented by $\dot{x}^{obs}$ and $\dot{P}_{b}^{obs}$)
in pulsar timing measurements might be caused by  geodetic
precession, an interesting gravitational effect we have been
seeking for.

\end{abstract}
 \draft 

\maketitle

\section{Introduction}
In the first gravitational two-body equation including
spins\cite{bo}, the precession velocity of the orbital angular
momentum vector, ${\bf L}$, was expressed as  around a vector,
combined by the spin angular momenta of the pulsar, ${\bf S}_1$,
and its companion ${\bf S}_2$. This velocity has been ignored,
since it is much smaller, 2 Post-Newtonian order (PPN), than the
precession velocity of the pulsar spin axis, 1.5PPN in binary
pulsar systems.

However, the small orbital precession velocity relative to the
vector combined by ${\bf S}_1$ and ${\bf S}_2$ doesn't mean that
it is also small relative to the line of sight (LOS). Then in the
study of observational effect of the orbital precession due to the
coupling of the spin induced quadrupole moment with the orbital
angular momentum \cite{sb,lai,wex2}, the precession of ${\bf L}$
is expressed as relative the total angular momentum vector, ${\bf
J}$, which is at rest to LOS (counting out the proper motion
effect); rather than the vector (${\bf S}_1$ and ${\bf S}_2$),
which are precessing relative to the LOS rapidly themselves.

Thus, there seems a contradictory between Barker and O'Connell's
equation and that of other authors mention above on which
direction ${\bf L}$ should precess around. Actually they can be
consistent in the following scenario\cite{sb,hs}. In which ${\bf
L}$, ${\bf S}_1$ and ${\bf S}_2$ all precess around ${\bf J}$
rapidly (1.5PPN), while the relative velocities of ${\bf L}$
relative to ${\bf S}_1$ and ${\bf S}_2$ are very small (2PPN).

The coupling of the spin induced quadrupole moment and the orbit
is 2\,PPN \cite{bo}, so the quadrupole moment of  the companion
star has to be very large to make the effect observable.
Therefore, this model is suitable for such binary pulsars, as
neutron star-main sequence star (NS-MS), and neutron star-black
hole (NS-BH)\cite{lai,wex2}.

Unlike NS-MS and NS-BH binaries, neutron star-white dwarf (NS-WD)
and neutron star-neutron star (NS-NS) binaries have much smaller
quadrupole moment, and the spin angular momenta of the two stars
can be comparable. To explain the observations of such binaries,
alternative model is necessary.

Apostolatos et al\cite{apo} and Kidder\cite{kid} studied the
precession of  ${\bf L}$ as relative to ${\bf J}$ in discussing
the modulation of gravitational wave by the spin-orbit (S-L)
coupling in coalescing binary systems (actually NS-BH). Similar to
the former quadrupole-orbit coupling, this S-L coupling only
considers the spin of the companion star with the orbit, and the
spin of the pulsar is ignored. The difference is that the former
quadrupole-orbit coupling corresponds to 2\,PPN, whereas the
latter S-L coupling corresponds to 1.5 PPN. Therefore, the latter
is easier to explain the measurements of NS-WD and NS-NS binaries,
which have smaller quadrupole moments and comparable spins.

This paper develops the equation of orbital precession of
Apostolatos et al and Kidder, making it applicable for general
binary pulsars. We point out that once a binary system has two
spins instead of one (the  masses of the binary system can be
different), the precession of the spin angular momenta of the two
stars, ${\bf S}_1$ and ${\bf S}_2$, lead to a variable ${\bf S}$
(${\bf S}\equiv{\bf S}_1+{\bf S}_2$), which in turn tilts the
orbital plane. Thus the additional motion of the orbital plane can
explain both $\dot{x}$ and $\dot{P}_{b}$ measured in NS-WDs, i.e.,
PSR~B1957$+$20\cite{aft} and PSR~J2051$-$0827\cite{do}, which have
been interpreted separately by different models\cite{wex2, as}.
Moreover, the derivatives, $\ddot{P}_{b}$, $\dddot{P}_{b}$, and
$\ddot{x}$ can also be naturally interpreted.

The geodetic precession induced $\dot{x}$ and $\dot{P}_{b}$
include an angle, $\lambda_{LJ}$ (the angle between ${\bf L}$ and
${\bf J}$), which represents the intensity of S-L coupling
($\lambda_{LJ}\approx S/L$). Comparing with the measured
$\dot{x}^{obs}$, the magnitude of S/L can be obtained. Since L can
be obtained from the binary parameters measured, we can thus
estimate the magnitude of ${\bf S}_1$ and ${\bf S}_2$, and in turn
the moment of inertia of the pulsar, through the measured pulsar
period. The obtained moment of inertia is about
$10^{44}$g~cm$^{2}$ and $10^{45}$g~cm$^{2}$, which is in the range
of strange stars and NSs.

On the other hand, the geodetic precession model is supported not
only by the secular variabilities measured, but also by
theoretical predictions of the moment of inertia of pulsars.

Section II introduces the orbital precession in special cases.
Section III discusses the orbital precession in general cases and
its relationship with the orbital precession in special cases.
 Section IV applies the
general orbital precession to two NS-WD binaries, explains the
significant secular variabilities measured and estimates the
moment of inertia of pulsars in the two binaries.
\section{orbital precession in special cases}
The motion of a binary system can be seen as the precession of
three vectors, the spin angular momenta of the pulsar and its
companion star, ${\bf S}_1$ and  ${\bf S}_2$,
and the orbital      
angular momentum $ {\bf L}$. The change of the orbital period due
to the gravitational radiation is 2.5\,PPN; whereas  the geodetic
precession corresponds to 1.5\,PPN. So the influence of
gravitational radiation on the motion of a binary system can be
ignored in the discussion of dynamics of a binary pulsar system.
Therefore, the total angular momentum, ${\bf J} = {\bf L} + {\bf
S}_1 + {\bf S}_2$, can be treated as invariable both in magnitude
and direction ($\dot{{\bf J}}=0$). With $\Omega_0$, the precession
rate of $\bf L$ around $\bf J$, the S-L coupling can be expressed
as\cite{bo}
\begin{equation}
\label{e1} {\bf \Omega}_0 \times {\bf L} = -{\bf \Omega}_1
 \times {\bf S}_1 - {\bf \Omega}_2 \times {\bf S}_2\,,
\end{equation}
where $\Omega_1$ and $\Omega_2$ represent the precessions of the
pulsar and its companion star, respectively. Ignoring those terms
that are over 2\,PPN, $\Omega_{1}$ and $\Omega_{2}$ can be written
as\cite{bo}
\begin{equation}
\label{e1aa}\Omega_{1}= \frac {L}{2r^{3}}(4+\frac{3m_{2}}{m_{1}})
\ , \ \ \Omega_{2}= \frac {L}{2r^{3}}(4+\frac{3m_{1}}{m_{2}})
 \ ,
\end{equation}
where $m_1$ and  $m_2$ are masses of the pulsar and the companion
star, respectively, and $r$ is the separation of  $m_1$ and $m_2$.
Notice $L\propto r^{1/2}$, $\Omega_1$ and $\Omega_2$ are 1.5\,PPN.

In the case of one spinning body, i.e., ${\bf S}_2=0$, ${\bf
S}={\bf S}_1$,  ${\bf L}$ and ${\bf S}$ precess about the fixed
vector ${\bf J}$ at the same rate with a precession frequency
approximately\cite{kid},
\begin{equation}
\label{e1aa4}\omega_{p}=\frac {|{\bf J}|}{2r^{3}}
(4+\frac{3m_{2}}{m_{1}}) \ .
\end{equation}
Eq($\ref{e1aa4}$) is also correct if the two bodies have equal
masses (${\bf \Omega}_1={\bf \Omega}_2$ in Eq($\ref{e1}$)).

Section III shows that the orbital precession in a general binary
pulsar system ($S_2\neq 0$, $S_2\neq 0$, $m_1\neq m_2$) will
automatically lead to the significant variability in $\dot{P}_{b}$
and $\dot{x}$.
\section{orbital precession and its effects in general cases}
Eq($\ref{e1}$) indicts that at any instance the variation of the
angular momenta ${\bf L}$ (left) equals the variation of the
angular momenta ${\bf S}_1$ and ${\bf S}_2$ (right). In magnitude,
the right-hand side of Eq($\ref{e1}$) is (recall ${\bf
S}\equiv{\bf S}_1+{\bf S}_2$)
\begin{equation}
\label{e1ax1} (\tau_{1})_{R}=\Omega_{2}S\sin\lambda_{LS}+
(\Omega_{1}-\Omega_{2})S^{\parallel}_{1}\sin \lambda_{LS_{1}}
 \ .
\end{equation}
$\bf L$ can react to the torque, $(\tau_1)_{\rm R}$, by precessing
around $\bf J$, with a very small opening angle of the precession
cone $\lambda_{LJ}\approx S/L$. Since the triangle (${\bf L}$,
${\bf J}$ and ${\bf S}$) constraint much be satisfied at any
instant. Thus the left-hand side of Eq($\ref{e1}$) can be written
as
\begin{equation}
\label{e1ax3}
 (\tau_{1})_{L}=|\Omega_{0}{\bf n}_{J}\times L{\bf
n}_{0}|=\Omega_{0}L\sin \lambda_{LJ} \ ,
\end{equation}
where ${\bf n}_J$ and ${\bf n}_0$ are unit vectors of $\bf J$ and
$\bf L$, respectively.
 By Eq($\ref{e1ax1}$) and Eq($\ref{e1ax3}$), we have the
 precession rate of ${\bf L}$ around ${\bf J}$,
\begin{equation}
\label{e1a} \Omega_{0}=\Omega_{2}\sin\lambda_{LS}+
(\Omega_{1}-\Omega_{2})\frac{S^{\parallel}_{1}}{S}\sin
\lambda_{LS_{1}}
 \ \,,
\end{equation}
where $S^{\parallel}_{1}=S_1\cos \eta_{SS1}$, which denotes the
component of $\bf{S}_1$ in the plane determined by $\bf{S}$ and
$\bf{J}$. Note that $L\sin \lambda_{LJ}\approx S$ is used in
Eq($\ref{e1a}$), since $S/L\ll 1$. The difference between
Eq($\ref{e1aa4}$) and Eq($\ref{e1a}$) is obviously due to
different number of spins included. Notice that the right-hand
side of Eq($\ref{e1a}$) can as well be written by replacing
subscribes 1 with 2 and 2 with 1.

Thus the orbital precession velocity of Eq($\ref{e1a}$),
$\Omega_0$, is derived via two assumptions: the conservation of
the total angular momentum; and the triangle constraint on the
precession cone of the orbit, $\lambda_{LJ}$.

Notice that $\Omega_0$ of Eq($\ref{e1a}$) is 1.5\,PPN. Which can
be absorbed by the advance of the precession of the periastron,
$\dot{\omega}$. The measured $\dot{\omega}$  is
given\cite{sb,wex2} by:
\begin{equation}
\label{e3i}\dot{\omega}^{obs}=\dot{\omega}^{GR}+\Omega_{0}\cos\lambda_{LJ}
\approx\dot{\omega}^{GR}+\Omega_{0}
\ \,.
\end{equation}
Since $\bf{S}_1$ and $\bf{S}_2$ precess with different velocities,
$\Omega_{1}$ and $\Omega_{2}$ respectively ($m_1\neq m_2$),
$\bf{S}$ varies both in magnitude and direction ($\bf{S}_1$,
$\bf{S}_2$ and $\bf{S}$ form a triangle). From the triangle of
$\bf{S}$, $\bf{L}$ and $\bf{J}$,  in reaction to the variation of
$\bf{S}$, $\bf{L}$ must vary in direction ($|{\bf L}|=$const),
which means the variation of $\lambda_{LJ}$ ($\bf{J}$ is
invariable). Therefore, by Eq($\ref{e3i}$), a variable
$\dot{\omega}^{obs}$ is expected.

The change of $\lambda_{LJ}$ also means the orbital plane tilts
back and forth, as shown in Fig~1, in turn both $\lambda_{LS}$ and
$\lambda_{JS}$ vary with time. Therefore, by Eq($\ref{e1a}$), the
derivative of the rate of orbital precession can be given by,
\begin{equation}
\label{en1} \dot{\Omega}_{0} =\Omega_{2}\Omega_{12}X_3X_4-
\Omega_{12}X_1(\Omega_{01}X_2+\Omega_{12}X_3) \ \,,
\end{equation}
where $\Omega_{12}=\Omega_{1}-\Omega_{2}$,
$\Omega_{01}=\Omega_{1}-\Omega_{0}$,
 $X_1=\frac{S^{\parallel}_{1}}{S}\sin \lambda_{LS_{1}}$,
 $X_2=\tan\eta_{ss1}$,
 $X_3=\frac{S_{V1}S_{V2}}{S^{2}}\frac{\sin\eta_{s1s2}}{\alpha\sin\lambda_{JS}}
 $, and
 $X_4=\frac{\cos^{2}\lambda_{LS}}{\sin \lambda_{LS}}$;
with $\alpha=\sin \lambda_{JS}+\frac{\cos
^{2}\lambda_{LS}}{\sin\lambda_{LS}}$, $S_{V1}=S_{1}\sin
\lambda_{JS1}$ and $S_{V2}=S_{2}\sin \lambda_{JS2}$ represent
components of $\bf{S}_{1}$ and $\bf{S}_{2}$ that are vertical to
$\bf{J}$.

 Note that $\Omega_1$ and
$\Omega_2$ are unchanged, and $\lambda_{LS_{\alpha}}$ are
unchanged, since they decay much slower than that of the
orbit\cite{apo}.
 The second derivative of $\Omega_{0}$ are given by,
$$ \ddot{\Omega}_{0}
=\Omega_{2}\Omega_{12}(\dot{X}_{3}X_{4}+X_{3}\dot{X}_{4})-
\Omega_{12}\dot{X}_{1}(\Omega_{01}X_{2}+\Omega_{12}X_{3})$$
\begin{equation}
\label{en1a}
-\Omega_{12}X_{1}(\Omega_{01}\dot{X}_{2}+\Omega_{12}\dot{X}_{3})
 \ \,,
\end{equation}
where $\dot{X}_{\alpha}$ is the first derivative of
${X}_{\alpha}$. $\dddot{\Omega}_{0}$ can be easily obtained from
Eq($\ref{en1a}$).

$\dot{\Omega}_{0}$, the derivative of $\Omega_{0}$, can be
absorbed by $\dot{P}_{b}$. The variation in the precession
velocity of the orbit results in a variation of orbital frequency
(${\nu}_{b}=2\pi/P_{b}$),
${\nu}^{\prime}_{b}-{\nu}_{b}=\dot{\Omega}_{0}\Delta t$, then we
have $\dot{\nu}_{b}=\dot{\Omega}_{0}$ and
$\ddot{\nu}_{b}=\ddot{\Omega}_{0}$, therefore,
\begin{equation}
\label{ez2} \dot{P}_{b}=-\frac{\dot{\Omega}_{0}P_{b}^{2}}{2\pi} \
\,, \ \ \ \
 \ddot{P}_{b}=-\frac{\ddot{\Omega}_{0}P_{b}^{2}}{2\pi}
-\frac{\dot{\Omega}_{0}P_{b}\dot{P}_{b}}{\pi}
\, .
\end{equation}
From Eq($\ref{en1}$) and Eq($\ref{ez2}$), we can see that the
contribution of the orbital precession to $\dot{P}_{b}$ is 1~PPN,
which is much larger than the contribution of GR to $\dot{P}_{b}$.
Since it depends on parameters, $X_1$, $X_2$, $X_3$, and
$\lambda_{LS}$ of Eq($\ref{en1}$), then it is also possible that
the geodetic precession induced $\dot{P}_{b}$ is comparable to, or
even smaller than the prediction of GR in special cases (special
combination of $\lambda_{LS}$, $\lambda_{LS1}$, $\eta_{SS1}$ and
$S_{1}/S$).

The effect of S-L coupling on secular evolution of the orbital
inclination, $i$, can be given as,
\begin{equation}
\label{e2}
 \cos i=\cos \lambda _{LJ} \cos I - \sin \lambda _{LJ} \sin I \cos \eta
 _{0} \ \,,
\end{equation}
where $I$ is the angle between the total angular momentum, ${\bf
J}$, and LOS, and $\eta _{0}=\Omega_{0}t+\eta _{i} $ ($\eta_{i} $
is the initial phase) is the phase of precession of ${\bf L}$.
Thus $i $ is also a function of time. By
 Eq($\ref{e2}$) we have,
\begin{equation}
\label{e3a} d i/d t =-\frac{\Omega_{0}S}{L}\sin \eta
_{0}+O(r^{-7/2}) \ \,.
\end{equation}
 Thus all
 parameters that are related with the orbital inclination will change
 with the geodetic precession. Hence the projected semi-major axis,
 $x=a_{1}\sin i/c$, will vary as the precession of the orbital
plane. By Eq($\ref{e3a}$), the derivatives of the projected
semi-major axis are
\begin{equation}
\label{e3c}\dot{x}=-x\Omega_{0}\sin \lambda_{LJ}\sin \eta_{0}\cot
i \ \,,
\end{equation}
\begin{equation}
\label{e3cc}\ddot{x}=-\dot{x}(\dot{\Omega}_{0}/\Omega_{0}+
\Omega_{0}\cot\eta_{0}) \ \,.
\end{equation}
As discussed between Eq($\ref{e3i}$) and Eq($\ref{en1}$), the
change of $S$ leads to the change of $\Omega_0$, and in turn
$\dot{P}_{b}$ and $\ddot{P}_{b}$. If there is only one spin, then
$S$ is a constant, thus there will be only a static addition
(${\Omega}_{0}=const$) to the apsidal motion, as shown in
Eq($\ref{e3i}$). In this case, the orbital precession is also
static, which will only cause a variation in $\dot{x}$, as shown
in Eq($\ref{e3c}$), whereas, $\dot{P}_{b}$ and $\ddot{P}_{b}$ are
not influenced at all ($\dot{\Omega}_{0}=0$ and
$\ddot{\Omega}_{0}=0$). This is why the quadrupole-orbit model
(also the one spin S-L coupling model) cannot explain
$\dot{P}_{b}$ and $\ddot{P}_{b}$ in PSR~J2051$-$0827 and
PSR~B1957$+$20.

\section {effects of  geodetic precession induced $\Omega_0$}
The S-L coupling induced $\Omega_0$ leads to the precession of the
orbital plane, then the orbital inclination, $i$, varies, as in
Eq($\ref{e3a}$), and in turn the projected semi-major axis,
$\dot{x}$, varies, as in Eq($\ref{e3c}$). Comparing the predicted
$\dot{x}$ with the observational $\dot{x}^{obs}$, we can constrain
the angle $\lambda_{LJ}\approx S/L$, and therefore the moment of
the inertia of the pulsar.
\subsection{PSR~J2051$-$0827}
The measured orbital period derivatives of
PSR~J2051$-$0827\cite{do} are list in Table~I, the derivatives of
the semi-major axis are $\dot{x}^{obs}=-23(3)\times 10^{-14}$ and
$\ddot{x}^{obs}=7(15)\times 10^{-22}$ s$^{-1}$.  The measured ones
are much larger than the corresponding predictions of GR,
$\dot{P}_{b}^{GR}=-3.8\times 10^{-14}$ and
$\dot{x}^{GR}=-1.3\times 10^{-19}$, as well as the proper motion
induced effects, $\dot{x}^{pm}=4(2)\times 10^{-17}$\cite{ko}.

With $m_1=1.40M_{\odot}$, $m_2=0.03M_{\odot}$ and
${P}_{b}=0.09911$d\cite{do}, we have the semi-major axis,
$a=(GM/\nu_{b}^{2})^{1/3}=7.08\times10^{10}$~cm, and orbital
angular momentum, $L=\nu_{b}\mu a^{2}(1-e^{2})^{1/2}=2.17\times
10^{50}$g~cm$^{2}$s$^{-1}$, with $\mu$ the reduced mass. From
Eq($\ref{e1aa}$), we have $\Omega_2=3.25\times 10^{-9}$s$^{-1}$,
and from Eq($\ref{e1a}$), $|\Omega_0|\approx\Omega_2\approx
3\times10^{-9}$s$^{-1}$. With $x=x^{obs}=0.045052$,
$|\dot{x}^{obs}|=2.3\times 10^{-13}$, Eq($\ref{e3c}$) becomes
\begin{equation}
\label{e2051a}
 0.045\times(3\times10^{-9})\sin \lambda_{LJ}|\sin\eta_{0}\cot i|=2.3\times 10^{-13}  \,.
\end{equation}
Thus $\sin\lambda_{LJ}|\sin\eta_{0}\cot i|\approx 2\times10^{-3}$.
  If $|\sin \eta_{0}\cot i| \approx 1\times 10^{-1}$, then $\lambda_{LJ}\approx
  2\times10^{-2}$, which means
 $S/L\approx 2\times10^{-2}$. With $L$ obtained above, we have $S_1\approx S\approx 4
\times10^{48}$g~cm$^{2}$s$^{-1}$. Having the measured pulsar
period, $P=4.50864$ms, the moment of inertia of the pulsar is
$I_{1}\approx 3\times10^{45}$g~cm$^{2}$.

\subsection{PSR~B1957$+$20}
The negative orbital period derivative changes to positive during
the observation of PSR~B1957$+$20\cite{aft}. The measured
derivatives of the orbital period are shown in Table~2. The
measured upper-limit to $\dot{x}$ is $\dot{x}^{obs}<3\times
10^{-14}$. The measured ones are also much larger than the
corresponding GR predictions, $\dot{P}_{b}^{GR}=-3.4\times
10^{-15}$ and $\dot{x}^{GR}=-6.1\times 10^{-21}$, as well as the
proper motion induced effect, $\dot{x}^{pm}=2.4\times
10^{-16}$\cite{ko}.

Similarly, with $m_1=1.40M_{\odot}$, $m_2=0.025M_{\odot}$ and
${P}_{b}=33001.91484$s\cite{aft}, we have
$a=1.74\times10^{11}$~cm, $L=2.82\times
10^{50}$g~cm$^{2}$s$^{-1}$, $\Omega_2=3.43\times10^{-10}$s$^{-1}$,
and in turn $|\Omega_0|\approx\Omega_2\approx
3\times10^{-10}$s$^{-1}$. From Eq($\ref{e3c}$), with
$x=x^{obs}=0.0892253$ and $|\dot{x}^{obs}|<3\times 10^{-14}$, we
have
\begin{equation}
\label{e1957a}
 0.089\times (3\times10^{-10})\sin \lambda_{LJ}|\sin
\eta_{0}\cot i|<3\times 10^{-14} \ \,,
\end{equation}
 which leads to ,
 $\sin\lambda_{LJ}|\sin\eta_{0}\cot i|<1\times10^{-3}$.
If $|\sin\eta_{0}\cot i|\approx1\times10^{-1}$, then
$\lambda_{LJ}<1\times10^{-2}$. Which means $S/L<1\times10^{-2}$,
and by the obtained $L$, we have $S_1\approx S<3
\times10^{48}$g~cm$^{2}$s$^{-1}$. Finally, with the measured
pulsar period, $P=1.607$ms, the moment of inertia of the pulsar is
$I_{1}<8\times10^{44}$g~cm$^{2}$ .

The simple estimation above indicates that PSR~B1957$+$20 has
similar moment of inertia as that of PSR~J2051$-$0827.

 In these
two estimations, the assumptions $|\Omega_0|\approx\Omega_2$ and
$|\sin\eta_{0}\cot i|\approx1\times10^{-1}$ are used. Consider the
deviation of these two assumptions from the corresponding true
values, the most likely moment of inertia from the observational
constraints is about $10^{44}$g~cm$^{2}$--$10^{45}$g~cm$^{2}$.

\section {effects of  geodetic precession induced
$\dot{\Omega}_0$, $\ddot{\Omega}_0$, $\dddot{\Omega}_0$}

From Eq($\ref{en1}$) and Eq($\ref{en1a}$), we can see that
$\dot{\Omega}_0$, $\ddot{\Omega}_0$ and $\dddot{\Omega}_0$ can
vary in wider and wider range, which can be written approximately
as (for the convenience of estimation),
\begin{equation}
\label{eom1} |\dot{\Omega}_{0}|\approx c_1\Omega_{2}^{2}  \ \,, \
\ |\ddot{\Omega}_{0}|\approx c_2\Omega_{2}^{3}  \ \,, \ \
|\dddot{\Omega}_{0}|\approx c_3\Omega_{2}^{4} \ \,,
\end{equation}
where $c_1$ represents the contribution of $X_{\alpha}$ to
$\dot{\Omega}_{0}$,  $c_2$ corresponds to the contribution of
$X_{\alpha}$ and $\dot{X}_{\alpha}$ to  $\ddot{\Omega}_{0}$, and
$c_3$ corresponds to the contribution of $X_{\alpha}$,
$\dot{X}_{\alpha}$ and $\ddot{X}_{\alpha}$ to
$\dddot{\Omega}_{0}$. Obviously $\dddot{\Omega}_{0}$ can vary in a
wider range than $\ddot{\Omega}_{0}$, and $\ddot{\Omega}_{0}$ can
vary in a wider range than $\dot{\Omega}_{0}$.

For PSR~J2051$-$0827, $\Omega_{2}^{2}\approx 9.8\times
10^{-18}$s$^{-2}$, $\Omega_{2}^{3}\approx 3.2\times
10^{-26}$s$^{-3}$ and also (equation not shown)
$\Omega_{2}^{4}\approx 1.0\times 10^{-34}$s$^{-4}$. Assuming
$c_1=c_2=c_3=0.2$, and by Eq($\ref{eom1}$) and Eq($\ref{ez2}$),
the derivatives of $P_{b}$ can be obtained, $|\dot{P}_{b}|\approx
2\times 10^{-11}$, $|\ddot{P}_{b}|\approx 8\times
10^{-20}$s$^{-1}$ and $|\dddot{P}_{b}|\approx 2\times
10^{-28}$s$^{-2}$. Which can be well consistent with the
corresponding observations, as shown in Table~I.

For PSR~B1957$+$20, the measured  derivatives of ${P}_{b}^{obs}$
is larger than that of PSR~J2051$-$0827. By assuming  $c_1=1$,
$c_2=1\times 10^{3}$ and $c_3=1\times10^{4}$, and through the same
treatment, we have $|\dot{P}_{b}|\approx 2\times 10^{-11}$,
$|\ddot{P}_{b}|\approx 7\times 10^{-18}$s$^{-1}$ and
$|\dddot{P}_{b}|\approx 2\times 10^{-26}$s$^{-2}$, which can be
consistent with the corresponding observations, as shown in Table
II.

It is likely that in PSR~B1957$+$20 the magnitude of $S_1$ and
$S_2$ are close, so that $S$ can be small, which leads to a large
${X}_{\alpha}$, and in turn large derivatives of ${P}_{b}$ as
shown in Eq($\ref{e1a}$) and Eq($\ref{en1}$)-Eq($\ref{ez2}$).

\section{discussion and conclusion}
Geodetic precession induced pulsar spin in binary pulsar PSR
B1913+16 has been studied through structure parameters of pulsar
profile\cite{kr,wt}. Actually the variation of pulsar spin axis
(former) and the  measured  secular variabilities discussed in
this paper are induced by the same physics underlying in binary
pulsar systems, S-L coupling. The former corresponds to the
reaction of the pulsar spin to the torque in the S-L coupling;
while the latter corresponds to reaction of the orbit to the same
torque.

By adding one more spin into the S-L coupling of Apostolatos et al
and Kidder\cite{apo,kid}, we establish the relationship among the
secular variability, geodetic precession and moment of inertia of
binary pulsars. The new model provides: (a) a unified model that
explain both $\dot{x}$ and $\dot{P}_{b}$, which has been
interpreted separately by different models, (b) a new method which
can extract the moment of inertia through pulsar timing
measurement, (c) a  new test of the geodetic precession which has
strong effects to pulsar timing.



\begin{table}
\begin{center}
\caption{Measured parameters compare with the geodetic precession
induced ones in PSR~J2051$-$0827}
\begin{tabular}{ll}
\hline \hline
 observations & predictions     \\ \hline
 $\dot{x}^{obs}
=-23(3)\times 10^{-14}$
  & $|\dot{x}|\approx 2\times 10^{-13} $ $^a$    \\

$({\ddot{x}}/{\dot{x}})^{obs}
 \lesssim-3.0\times 10^{-9}$s$^{-1}$
  & $|{\ddot{x}}/{\dot{x}}|\approx 3\times 10^{-9}$s$^{-1}$ $^{b}$    \\

$\dot{P}_{b}^{obs}=-15.5(8)\times 10^{-12}$ &
$|\dot{P}_{b}|=|\frac{\dot{\Omega}_{0}P_{b}^{2}}{2\pi}|\approx
2\times
10^{-11}$  \\

$\ddot{P}_{b}^{obs}=2.1(3)\times 10^{-20}$s$^{-1}$ &
$|\ddot{P}_{b}|\approx |\frac{\ddot{\Omega}_{0}P_{b}^{2}}{2\pi}|
\approx8\times 10^{-20}$s$^{-1}$  \\

$\dddot{P}_{b}^{obs}=3.6(6)\times 10^{-28}$s$^{-2}$ &
$|\dddot{P}_{b}|\approx |\frac{\dddot{\Omega}_{0}P_{b}^{2}}{2\pi}|
\approx 2\times 10^{-28}$s$^{-2}$
\\ \hline \hline
\end{tabular}
\end{center}
{\small We assume $c_1=c_2=c_3=0.2$ in Eq($\ref{eom1}$). $^a$
$|\dot{x}|$ is given by Eq($\ref{e2051a}$), and $^b$
$\ddot{x}/\dot{x}$ is given by Eq($\ref{e3cc}$).
 }
\end{table}

\begin{table}
\begin{center}
\caption{Measured parameters compare with the geodetic precession
induced ones in PSR~B1957$+$20}
\begin{tabular}{ll}
\hline \hline
 observations & predictions     \\ \hline
 $|\dot{x}|^{obs}
<3\times 10^{-14}$
  & $|\dot{x}|\lesssim 3\times 10^{-14} $ $^a$    \\

$\dot{P}_{b}^{obs}=1.47(8)\times 10^{-11}$ & $|\dot{P}_{b}|=
|\frac{\dot{\Omega}_{0}P_{b}^{2}}{2\pi}| \approx 2\times
10^{-11}$  \\

$\ddot{P}_{b}^{obs}=1.43(8)\times 10^{-18}$s$^{-1}$ &
$|\ddot{P}_{b}|\approx |\frac{\ddot{\Omega}_{0}P_{b}^{2}}{2\pi}|
\approx7\times 10^{-18}$s$^{-1}$  \\

$|\dddot{P}_{b}|^{obs}<3\times 10^{-26}$s$^{-2}$ &
$|\dddot{P}_{b}|\approx |\frac{\dddot{\Omega}_{0}P_{b}^{2}}{2\pi}|
 \approx 2\times
10^{-26}$s$^{-2}$
\\ \hline \hline
\end{tabular}
\end{center}
{\small We assume $c_1=1$, $c_2=1\times 10^{3}$, and
$c_3=1\times10^{4}$ in Eq($\ref{eom1}$). $^a$ $|\dot{x}|$ is given
by Eq($\ref{e1957a}$).
 }
\end{table}

\clearpage

\begin{figure}[t]

\begin{center}
\includegraphics[87,87][700,700]{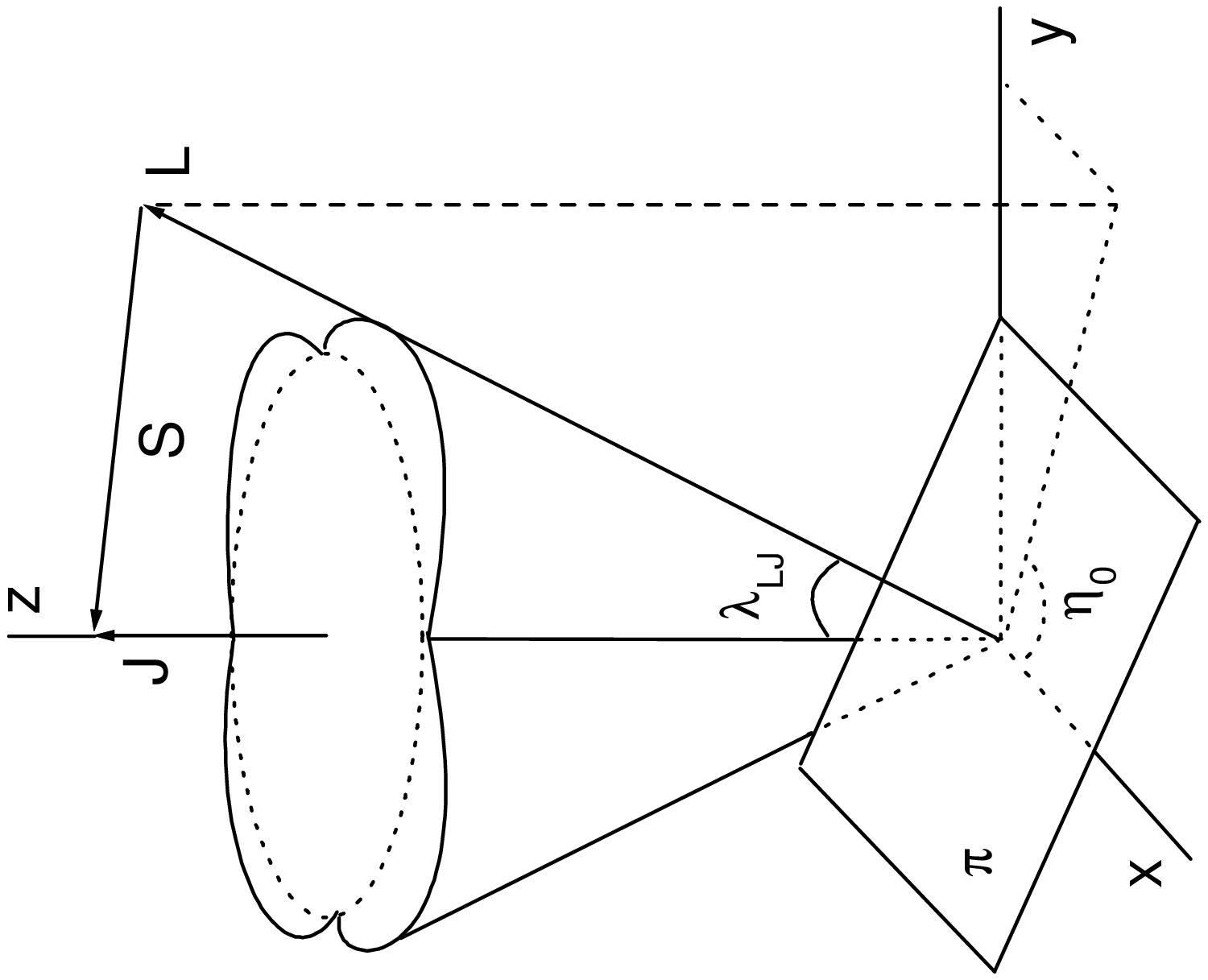}

\end{center}
\caption{The orbital plane, $\pi$, tilts back and forth (
corresponding to ${\bf L}$ moves along the curves). ${\bf L}$,
${\bf S}$ and ${\bf J}$ form a triangle, in which ${\bf J}$ is
invariable both in magnitude and direction; ${\bf L}$ varies in
direction only; and ${\bf S}$ varies both in  magnitude and
direction}
\end{figure}

\end{document}